\documentclass{IEEEtran}
\usepackage{cite}
\usepackage{multirow}
\usepackage{amsmath,amssymb,amsfonts}
\usepackage[ruled, vlined, linesnumbered]{algorithm2e}
\usepackage{graphicx}
\usepackage{textcomp}

\usepackage{grffile}
\usepackage{epstopdf}
\usepackage{mathptmx}      
\usepackage{float}
\usepackage{enumitem}
\usepackage{lipsum}
\usepackage{grffile}
\usepackage{latexsym}

\def\BibTeX{{\rm B\kern-.05em{\sc i\kern-.025em b}\kern-.08em
    T\kern-.1667em\lower.7ex\hbox{E}\kern-.125emX}}
\begin{document}
\title{An Efficient Hybrid Beamforming Design for Massive MIMO Receive Systems via SINR Maximization Based on an Improved Bat Algorithm}
\author{Mohammed A. Almagboul, Feng Shu, Yaolu Qin, Xiaobo Zhou, Jin Wang, Yuwen Qian, and Kingsley Jun Zou
\thanks{}
\thanks{Mohammed A. Almagboul, Feng Shu, Yaolu Qin, Xiaobo Zhou, Jin Wang, Yuwen Qian, and Kingsley Jun Zou
are with School of Electronic and Optical Engineering, Nanjing University of
Science and Technology, Nanjing, 210094, China.}
\thanks{Mohammed A. Almagboul is also with the Electronic Engineering Department
at Sudan Technological University, Khartoum, Sudan, and the Communication Engineering
Department, AlMughtaribeen University, Khartoum, Sudan.}
\thanks{Feng Shu is also with the College of Computer and Information Sciences,
Fujian Agriculture and Forestry University, Fuzhou 350002, China, and the
College of Physics and Information, Fuzhou University, Fuzhou 350116,
China.}}

\maketitle
\begin{abstract}
Hybrid analog and digital (HAD) beamforming has been recently receiving considerable deserved attention for a practical implementation on the large-scale antenna systems. As compared to full digital beamforming, partial-connected HAD beamforming can significantly reduce the hardware cost, complexity, and power consumption. In this paper, in order to mitigate the jamming along with lowering the hardware complexity and cost by reducing the number of RF chains needed, a novel hybrid analog and digital receive beamformer based on an improved bat algorithm (I-BA) and the phase-only is proposed. Our proposed beamformer is compared with robust adaptive beamformers (RABs) methods proposed by us, which are considered in the digital beamforming part. The evolutionary optimization algorithm is proposed since most of the RAB methods are sensitive to the DOA mismatch, and depending on the complex weights, resulting in an expensive receiver. In the analog part, analog phase alignment by linear searching (APALS) with a sufficiently fine grid of points is employed to optimize the analog beamformer matrix. The performance of the proposed I-BA is revealed using MATLAB simulation and compared with BA, and Particle swarm optimization (PSO) algorithms, which shows a better performance in terms of convergence speed, stability, and the ability to jump from the local minima.
\end{abstract}


\section{Introduction}
\label{sec:introduction}
\begin{figure*}[!t]
\centering{\includegraphics[width= 13cm,height=5.5cm]{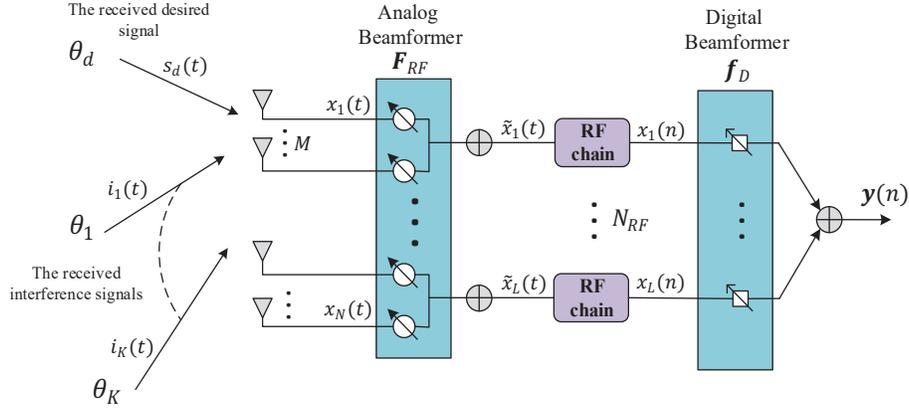}}  
\caption{Partial-connected hybrid analog and digital BF structure at the receiver\label{fig1}}
\end{figure*}

\IEEEPARstart{}{}Although the initial applications of adaptive beamforming were in military areas such as sonar and radar, its use in civilian applications like mobile communications, ultrasonics, and seismology also gained great popularity today [1]. Adaptive beamforming has well-known advantages like enhancing the system capacity and reduce the interference by steering the antenna array main beam pattern toward the desired signal, while steer the nulls toward the interference signals directions; this can be accomplished by constantly updating the total beamforming weights in a hybrid system. Due to the multiplicity and diversity of the sources of interference resulting from the wide spread of wireless devices, the employment of hybrid adaptive beamforming in smart transportation and unmanned aerial vehicles (UAVs) will be one of the most important techniques that can be used to avoid the risk of the undesired interference and jamming signals [2], whereas, with the potential widespread use of UAVs in the near future in many civilian applications, they face a highly critical threat using a relatively easy method i.e., drone jammers, particularly when used for goods delivery, where criminals can use jammers to obtain goods or obtain the UAV itself. On the other hand, reducing the UAV payload and power consumption are considered one of the most important design goals. Therefore, the use of the hybrid system is very desirable, where it allows us to reduce the number of RF chains in the system compared to a fully digital beamforming design. There are a number of algorithms and ways to design the adaptive beamforming; an algorithm that has high convergence speed, low computation, low complexity, and better performance is surely needed.

The immense hardware cost and power consumption for massive multiple-input multiple-output (MIMO) systems due to the large number of RF chains required for a single receiver unit is considered one of the most design challenges. Particularly, in the classic massive MIMO system each antenna is connected to one RF chain, resulting in a remarkably high hardware cost, complexity, and power consumption, where the RF chain includes power-avid components like amplifiers and analog-to-digital converter (ADC) [3]. Consequently, the use of the hybrid analog and digital (HAD) beamforming received great interest from researchers, due to promising practical implementation in the massive MIMO systems for 5G communications [4], where it requires fewer numbers of RF chains as compared to a fully digital beamforming design [5, 6]. Moreover, HAD beamforming combines the accuracy and speedy features of digital beamforming which compensate the reduction of RF chains, and the inexpensive characteristic of analog beamforming [7, 8].

Nature-inspired metaheuristic optimization algorithms are widely used recently to cope the restrictions of the conventional adaptive optimization algorithms based on error-derivative methods in terms of inflexibility, getting stuck in a local minima, and the need of accurate knowledge of directions of arrival (DOAs) of jamming signals, like linearly constrained minimum variance (LCMV), and minimum variance distortionless response (MVDR) [7, 9, 10]. The application of these metaheuristic algorithms in the array pattern synthesis in order to provide SLL minimization and steering the nulls in the desired interference directions has received great interest from researchers recently due to its flexibility, and being able to deal with non-convex and non-differentiable optimization problems [9-13]. Genetic algorithm (GA) [14] is one of the ancient and well-known nature-inspired metaheuristic techniques, where has been used early for synthesizing pattern of antenna arrays [15]. Particle swarm optimization (PSO) [16] is another widely known evolutionary algorithm; It is faster, efficient, easier to implement, and has a capability of solving linear and nonlinear optimization problems. PSO is used extensively for the designing of antenna arrays [17, 18]. A lot of other Meta-heuristic optimization algorithms also successfully employed in antenna arrays synthesizing applications [12, 19-21], such as the Ant-Lion optimization (ALO) technique introduced in [22], Grey Wolf Optimizer (GWO) [23], The cat swarm optimization (CSO) [24], ant colony optimization (ACO) [25], Invasive Weed Optimization (IWO) [26], simulated annealing (SA) [27], Whale Optimization Algorithm (WOA) [28], and others.

The Bat Algorithm (BA) is an evolutionary swarm intelligence algorithm initiated by Yang in 2010 [29], inspired from the nature behavior of bats, which use echolocation by changing pulse rates of emission and loudness to detect prey and avert obstacles. A number of researchers have used Bat algorithm for linear antenna array (LAA) to steer nulls and minimize sidelobe level (SLL). Tong and Truong [10] proved that the Bat algorithm can outperform the accelerated particle swarm optimization (APSO) and GA in terms of adaptive null-steering, sidelobe suppression, and computation time in array pattern nulling synthesis using phase-only control. In [30] again they utilize BA in order to minimize SLL and to place nulls in the desired directions using amplitude-only control, and proved that the BA based beamformer is more effective and faster as compared to GA and APSO. In [31] BA based beamformer using complex weight (amplitude and phase) compared with APSO shows faster convergence and higher efficiency. However, all above researches and a lot of other researches using different nature-inspired optimization algorithms mainly concentrate on the full digital beamformer instead of hybrid analog and digital system. To best of our knowledge, very few employments of these algorithms in hybrid analog and digital beamformer were carried out [32-36]. In [34, 35] the authors proposed a phase-only hybrid analog and digital beamforming based on GA with the aim of minimizing the transmit power under signal-to-interference-plus-noise ratio (SINR) constraints. Based on the Particle Swarm Ant Colony Optimization (PSACO) algorithm, in [36] a partial-connected hybrid precoding structure for wideband Massive MIMO systems is proposed in order to realize excellent energy and spectral efficiency. In [33] a joint precoding in the multiuser MIMO system with the objective of maximizing the capacity is carried out with a genetic algorithm (GA). Furthermore, the authors in [32] proposed two hybrid digital and analog beamformers based on PSO and manifold optimization (MO) in order of maximizing capacity. However, all these research activities mostly focusing on a transmitter not receiver. The authors in [37] proposed a transmitter/receiver based on OFDM, random subcarrier selection (RSCS), and directional modulation (DM) for secure messages transmission. In general, the proposed beamformers concentrating mainly on the transmitter, and the objective issues such as capacity maximization, secrecy rate maximization [38], maximizing signal-to-leakage-and-noise ratio (Max-SLNR) per user [39, 40], and transmit power minimization, while often neglecting the robustness. Some research activities provide improvements on BA to avoid the weaknesses of the algorithm being trapped in the local minima or yielded unstable results [41-43].
This paper focuses on an efficient interference suppression at the receiver using partial-connected HAD beamforming structure via maximizing SINR in the case of the presence and absence of DOA mismatches. The main contributions of this paper are summarized as follows:
\begin{enumerate}
\item Present an Improved-BA and analyze its special properties and excellent features in terms of adaptive beamforming. The main distinctions between the Improved-BA and the BA are the bats$'$ compensation for Doppler effects in echoes, and the possibility of selecting different habitats; this makes the algorithm further imitating the bats$'$ behaviors and thus improves the stability and efficiency.
\item An efficient partial-connected HAD receive beamformer based on Improved-BA is proposed, which highly reduces the cost and complexity, maximizing SINR and steering the nulls in the directions of interferences.
\item To guarantee high efficiency and robustness of our design, we first optimize the digital baseband beamformers' vector by means of closed-form solution using robust adaptive beamforming methods, namely diagonal-loading (DL) technique [44], and spatial matched filter (SMF), which control both amplitude and phase (complex weights). Moreover, we propose an efficient I-BA optimization algorithm to optimize the digital beamformer vector by controlling only the phase, resulting in an inexpensive and easy-to-implement receiver. Thereafter, analog beamformer's weights vectors are optimized through analog phase alignment (APA) by linear search (LS) using a sufficiently fine grid of points in the whole DOAs range. The performance of the proposed algorithm is compared with BA, and PSO evolutionary algorithms, which shows better ability to jump from the local minima, faster convergence speed, and relatively high stability.  Furthermore, the proposed hybrid beamformer I-BA-APALS is compared with other proposed hybrid beamformers DL-APALS and SMF-APALS, in addition to the conventional fully digital standard capon beamformer (SCB), and DL techniques, in terms of SINR obtained, robustness, and nulls depth for a given SNR of the desired and interference signals, which showed better performance compared to others.
\end{enumerate}
The rest of this work is structured as follows: in the following, our system model of partial-connected hybrid analog and digital beamformer is defined. In section III the problem formulated. Thereafter, basic BA and proposed Improved-BA are described, and the results discussed in Section IV, and V respectively. Finally, conclusions are given in Section VI.

\textbf{Notation}: Capital $X$, boldface small $\textbf{x}$, and small x letters are used to represent matrices, vectors, and scalars, respectively. Notations ${{(\cdot )}^{T}}$  and ${{(\cdot )}^{H}}$  denote transpose and conjugate transpose of a matrix, respectively. $\left| ~x~ \right|$ denotes the magnitude of a complex number $x$, while $\mathbb{E}\left\{ \cdot  \right\}$ denotes the expectation operation.

\section{System model}\label{sec2}
Assume an N-elements partial-connected HAD beamformer structure at the receiver. Its block diagram is shown in Fig. 1, which used to reduce the hardware cost and energy consumption with somewhat less performance. In this structure, the receiver is chosen to be equipped with $N$ isotropic antennas divided into $L$ subsets of antenna arrays, and each subset contains $M$ antenna elements, where the number of RF chains ${{N}_{RF}}$ chosen to be less than the number of antenna elements ${{N}_{RF}}\le N$. Each subset of antenna array elements connected to only one RF chain. The antenna elements are followed by a phase shifters that feed the RF chains. In Fig. 1, hybrid BF receives one desired signal ${{s}_{d}}\left( t \right){{e}^{j2\pi {{f}_{c}}t}}$ with an angle of arrival (AOA) ${{\theta }_{d}}$, and $K$ interference signals ${{i}_{k}}\left( t \right){{e}^{j2\pi {{f}_{c}}t}}$ with different angles of arrival ${{\theta }_{k}}$ $(k=1,...,K)$. The received signals ${{x}_{m}}\left( t \right)$ of the $lth$ sub-array at the input of each $mth$ element $(m=1,...,M)$ includes the desired narrow band signal, the interference narrow band signals, and an additive Gaussian noise $v\left( t \right)$ with zero mean and variance $\sigma _{n}^{2}$. Therefore, the $lth$ sub-array output ${{\tilde{x}}_{l}}\left( t \right)$ can be represented as follows,
\begin{align}\label{jkl}
\tilde{x}_{l}(t)&=\sum\limits_{m=0}^{M-1}s_d(t)e^{j2\pi {{f}_{c}}\left( t-\left( {\tau_d}-\frac{\left( \left( l-1 \right)M+m-1 \right)d}{c}sin{{\theta }_{d}} \right)-\frac{{{\alpha }_{l,m}}}{2\pi {{f}_{c}}} \right)}+\nonumber\\
&\sum\limits_{k=1}^{K}\sum\limits_{m=0}^{M-1}i_k(t)e^{j2\pi {{f}_{c}}\left( t-\left( {\tau_k}-\frac{\left( \left(l-1 \right)M+m-1 \right)d}{c}sin{{\theta }_{k}} \right)-\frac{{{\alpha }_{l,m}}}{2\pi {{f}_{c}}} \right)}+\nonumber\\
&v_{l}(t)
\end{align}
where, $d$ is the spacing between adjacent antenna array elements assumed throughout this paper to be $0.5\lambda $, $\tau_d$ and $\tau_k$ are the propagation delays from the desired signal emitter and kth interference signal emitter, respectively, to a reference point which is assumed to be the first element on the array, and $c$ is the speed of light. Eq. (1) includes the phase difference $\alpha_{l,m}$ for the $mth$ phase shifter of the analog beamformer in the $lth$ sub-array. After the analog beamformer, the signal $\tilde{x}_{l}(t)$ passes through the $L$ RF chains which include ADCs and down converters, resulting the following baseband signal in a matrix-vector notation for all $L$ subsets,
\begin{equation}\label{xy}
\textbf{x}(n)=F_{RF}^{H}A\textbf{s}(n)+\textbf{v}(n),
\end{equation}
where
\begin{align}
&{{F}_{RF}}=diag\left( {{\textbf{f}}_{1}},...,~~{{\textbf{f}}_{l~}},.~.~.,{{\textbf{f}}_{L}} \right) \nonumber \\
&{{\textbf{f}}_{l~}}=~\frac{1}{\sqrt{M}}{{\left[ \exp \left( j{{\alpha }_{1,l}} \right),~\exp \left( j{{\alpha }_{2,l}} \right),~.~.~.,~\exp \left( j{{\alpha }_{M,l}} \right)~ \right]}^{T}}
\end{align}
A matrix ${F}_{RF}$ is the $N\times L$ phase shift matrix, $\textbf{s}(n)=[{s}_{d}(n), \\
\text{ }\!\!~\!\!\text{ }{{i}_{1}}\left( n \right),\text{ }\!\!~\!\!\text{ }\ldots ,\text{ }\!\!~\!\!\text{ }{i}_{K}(n)]^{T}$, $\textbf{v}(n)=[{v}_{1}(n),\text{ }\!\!~\!\!\text{ }{{v}_{2}}\left( n \right),\text{ }\!\!~\!\!\text{ }\ldots ,\text{ }\!\!~\!\!\text{ }{v}_{L}(n)]^{T}$ is an additive white Gaussian noise (AWGN), and $A$ is $N\times (K+1)$ matrix of steering vectors $\textbf{a}(\theta)$,
\begin{equation}
A=\left[ \text{\textbf{a}}\left( {{\theta }_{d}} \right),\text{ }\!\!~\!\!\text{ \textbf{a}}\left( {{\theta }_{1}} \right),\text{ }\!\!~\!\!\text{  }\!\!~\!\!\text{ }.\text{ }\!\!~\!\!\text{  }\!\!~\!\!\text{ }.\text{ }\!\!~\!\!\text{  }\!\!~\!\!\text{ }.\text{ }\!\!~\!\!\text{  }\!\!~\!\!\text{ },\text{ }\!\!~\!\!\text{ \textbf{a}}\left( {{\theta }_{K}} \right) \right],
\end{equation}
where the column vector $\textbf{a}(\theta)$ is called an array manifold which can be given by,
\begin{equation}
\mathbf{a}\left( \text{ }\!\!\theta\!\!\text{ } \right)={{\left[ 1,\text{ }\!\!~\!\!\text{ }{{e}^{j\text{ }\!\!\pi\!\!\text{ }sin\theta }},.\text{ }\!\!~\!\!\text{ }.\text{ }\!\!~\!\!\text{ }.\text{ }\!\!~\!\!\text{  }\!\!~\!\!\text{ },\text{ }\!\!~\!\!\text{ }{{e}^{j\pi \left( N-1 \right)sin\theta }} \right]}^{T}}
\end{equation}
After the digital beamformer, ${{\textbf{f}}_{D}}\in {{\mathbb{C}}^{{{N}_{RF}}\times 1}}$, Eq. (3) becomes,
\begin{equation}
\textbf{y}\left( n \right)=\text{ }\!\!~\!\!\text{ }\textbf{f}_{D}^{H}F_{RF}^{H}A\textbf{s}\left( n \right)+\textbf{f}_{D}^{H}\textbf{v}\left( n \right)
\end{equation}
Through the digital beamforming vector $\textbf{f}_{D}$ we can control the amplitude, phase, or both.
\begin{equation}
{{\textbf{f}}_{D}}=\left[ {{a}_{1}}{{e}^{j{{\alpha }_{1}}}},\text{ }\!\!~\!\!\text{ }{{a}_{2}}{{e}^{j{{\alpha }_{2}}}},\text{ }\!\!~\!\!\text{ }.\text{ }\!\!~\!\!\text{ }.\text{ }\!\!~\!\!\text{ }.\text{ }\!\!~\!\!\text{ },\text{ }\!\!~\!\!\text{ }{{a}_{L}}{{e}^{j{{\alpha }_{L}}}} \right]^{T}
\end{equation}

\section{Problem formulation for SINR maximization}\label{sec3}
The input observation vector $\textbf{x}(t)$, can be formulated for all sub-arrays as,
\begin{align}
\textbf{x}(t)&=\mathbf{a}({\theta }_{d}){{s}_{d}}(t)~+~\underset{k=1}{\overset{K}{\mathop \sum }}\,\mathbf{a}\left( {{\theta }_{k}} \right){{i}_{k}}(t)+\textbf{v}(t)\nonumber\\
&=\text{ }\!\!~\!\!\text{  }\!\!~\!\!\text{ }\mathbf{a}({\theta }_{d}){{s}_{d}}(t)~+\text{ }\!\!~\!\!\text{ }{{A}_{i}}\textbf{i}(t)~+\textbf{v}(t)\nonumber\\
&=~~{{\textbf{x}}_{s}}(t)~+~{{\textbf{x}}_{i}}(t)~+~\textbf{v}(t),
\end{align}

where ${{A}_{i}}$ matrix constitutes of all steering vectors of interference signals, and $\textbf{i}(t)=[{i}_{1}(t),\text{ }\!\!~\!\!\text{ }{{i}_{2}}\left( t \right),\text{ }\!\!~\!\!\text{ }\ldots ,\text{ }\!\!~\!\!\text{ }{i}_{K}(t)]^{T}$. In this work, we assume that all the signals are zero mean, and independent.  Multiplying these signals by analog and digital beamformer weights and adding them together resulting in the following output,
\begin{equation}
\textbf{y}(n)=\textbf{f}_{D}^{H}F_{_{RF}}^{H}({{\textbf{x}}_{s}}(n)+{{\textbf{x}}_{i}}(n))+\textbf{f}_{D}^{H}\textbf{v}(n)
\end{equation}
The correlation matrix estimation can be composed from the signal and noise samples at $n$ time intervals, which is given by,
\begin{align}
{P}_{T}=~&{\mathbb{E}}\left\{ {{\textbf{s}}_{d}}(n)\textbf{s}_{d}^{H}(n) \right\}+\underset{k=1}{\overset{K}{\mathop \sum }}\,{\mathbb{E}}\left\{ {{\textbf{i}}_{k}}\left( n \right)\textbf{i}_{k}^{H}\left( n \right) \right\}+\nonumber\\
&{\mathbb{E}}\left\{ \textbf{v}\left( n \right){{\textbf{v}}^{H}}\left( n \right) \right\}\nonumber\\
=~&{{P}_{d}}+{{P}_{i}}+{{P}_{v}},
\end{align}
where ${P}_{d}$ represents the desired signal self-correlation matrix, while ${P}_{i}$, and ${P}_{v}$ represent the undesired interferences plus noise input signals self-correlation matrices respectively. The SINR is given by dividing the power of the desired signal by the sum of powers of all interference and noise signals. Thus, the hybrid system output power for the signal of interest can be given by,
\begin{align}
\sigma _{s}^{2}&=\text{ }\!\!~\!\!\text{ }{\mathbb{E}}\left\{ {{\left| \textbf{f}_{D}^{H}F_{RF}^{H}{{\textbf{x}}_{s}}\left( n \right) \right|}^{2}} \right\}\nonumber\\
&={\mathbb{E}}\left\{ {{\left| \textbf{f}_{D}^{H}F_{RF}^{H}\mathbf{a}\left( {{\theta }_{d}} \right){{\textbf{s}}_{d}}\left( n \right) \right|}^{2}} \right\}\nonumber\\
&={{P}_{d}}~\textbf{f}_{D}^{H}F_{RF}^{H}\textbf{a}\left( {{\theta }_{d}} \right){{\textbf{a}}^{H}}\left( {{\theta }_{d}} \right){{F}_{RF}}{{\textbf{f}}_{D}}
\end{align}
In the same context, we can derive the hybrid system output power for the unwanted signals as follows,
\begin{align}
\sigma _{i}^{2}&=\textbf{f}_{D}^{H}F_{RF}^{H}{{A}_{i}}{{P}_{i}}A_{i}^{H}{{F}_{RF}}{{\textbf{f}}_{D}} \\
\sigma _{v}^{2}&={{P}_{v}}~\textbf{f}_{D}^{H}{{\textbf{f}}_{D}}
\end{align}
Therefore, the SINR is defined as,
\begin{equation}
SINR=\frac{\sigma _{s}^{2}}{\sigma _{i}^{2}+\sigma _{v}^{2}~}~=\frac{{{P}_{d}}~\textbf{f}_{D}^{H}F_{RF}^{H}\textbf{a}\left( {{\theta }_{d}} \right){{\textbf{a}}^{H}}\left( {{\theta }_{d}} \right){{F}_{RF}}{{\textbf{f}}_{D}}}{~\textbf{f}_{D}^{H}F_{RF}^{H}{{A}_{i}}{{P}_{i}}A_{i}^{H}{{F}_{RF}}{{\textbf{f}}_{D}}+{{P}_{v}}~\textbf{f}_{D}^{H}{{\textbf{f}}_{D}}}
\end{equation}
Since ${{v}_{l}}\left( n \right)$ is an uncorrelated noise signal with zero mean and variance $\sigma ^{2}$, we get ${{P}_{v}}={\sigma ^{2}}I$. Without loss of generality, assume ${{P}_{i}}=I$, where $I$ is an identity matrix with appropriate size, Eq. (14) can be rewritten as follows,
\begin{equation}
SINR=\frac{{{P}_{d}}{{\left| \textbf{f}_{D}^{H}F_{RF}^{H}\textbf{a}\left( {{\theta }_{d}} \right) \right|}^{2}}}{~\textbf{f}_{D}^{H}F_{RF}^{H}{{A}_{i}}A_{i}^{H}{{F}_{RF}}{{\textbf{f}}_{D}}+{{\sigma }^{2}}~\textbf{f}_{D}^{H}{{\textbf{f}}_{D}}}
\end{equation}
Our goal is to maximize SINR. However, evolutionary algorithms usually looking for the minima, thus, the cost function (CF) can be given by the inverse of SINR as follows,
\begin{align}
minimize\,\,\,~CF=~~&\frac{~\textbf{f}_{D}^{H}F_{RF}^{H}{{A}_{i}}A_{i}^{H}{{F}_{RF}}{{\textbf{f}}_{D}}+~{{\sigma }^{2}}~\textbf{f}_{D}^{H}{{\textbf{f}}_{D}}}{{{P}_{d}}{{\left| \textbf{f}_{D}^{H}F_{RF}^{H}\textbf{a}\left( {{\theta }_{d}} \right) \right|}^{2}}}\nonumber\\
s.t.~~~~~~~~~~~~~~~~&\textbf{f}_{D}^{H}F_{RF}^{H}\textbf{a}\left( {{\theta }_{d}} \right)=1
\end{align}
In order of perfect extraction of the desired signal the constraint $\textbf{f}_{D}^{H}F_{RF}^{H}\textbf{a}\left( {{\theta }_{d}} \right)=1$ must be realized. Solving for digital beamformer: maximizing SINR can also be done by solving the following equivalent quadratic optimization problem,
\begin{align}
\underset{{{\textbf{f}}_{D}}}{\mathop{\min \text{ }\!\!~\!\!\text{ }}}\,~~~~~&\textbf{f}_{D}^{H}F_{RF}^{H}{{R}_{i+n}}{{F}_{RF}}{{\textbf{f}}_{D}}\nonumber\\
s.t.~~~~~~~&\textbf{f}_{D}^{H}F_{RF}^{H}\textbf{a}\left( {{\theta }_{d}} \right)=1,
\end{align}
where
\begin{equation}
{{R}_{i+n}}=~~\underset{k=1}{\overset{K}{\mathop \sum }}\,{{P}_{k}}\textbf{a}\left( {{\theta }_{k}} \right){{\textbf{\textbf{a}}}^{H}}\left( {{\theta }_{k}} \right)+{{P}_{v}}
\end{equation}

\subsection{Design of the  digital weight vector}
If the actual covariance matrix, ${{R}_{i+n}}$, are well known, Problem (17) can be solved using Lagrange's multiplier technique as,
\begin{equation}
{{\textbf{f}}_{D}}=~\frac{R_{i+n}^{-1}F_{RF}^{H}\textbf{a}\left( {{\theta }_{d}} \right)}{{{(F_{RF}^{H}\textbf{a}\left( {{\theta }_{d}} \right))}^{H}}R_{i+n}^{-1}F_{RF}^{H}\textbf{a}\left( {{\theta }_{d}} \right)}
\end{equation}
Here, to calculate ${{\textbf{f}}_{D}}$ we use the initial value of $F_{RF}$, that makes the array main beam steers towards the direction of the desired signal. Since $\textbf{f}_{D}^{H}F_{RF}^{H}\textbf{a}\left( {{\theta }_{d}} \right)=1$, the optimization problem (17) is equivalent to the following one,
\begin{align}
\underset{{{\textbf{f}}_{D}}}{\mathop{\min \text{ }\!\!~\!\!\text{ }}}\,~~~~~~&\textbf{f}_{D}^{H}F_{RF}^{H}{{R}_{i+n}}{{F}_{RF}}{{\textbf{f}}_{D}}+~{{P}_{d}}{{\left| \textbf{f}_{D}^{H}F_{RF}^{H}\textbf{a}\left( {{\theta }_{d}} \right) \right|}^{2}}\nonumber\\
s.t.~~~~~~~&\textbf{f}_{D}^{H}F_{RF}^{H}\textbf{a}\left( {{\theta }_{d}} \right)=1
\end{align}
This because the second term in the optimization problem is constant, so, it will not have an effect on the problem solution. Problem (20) can also be equivalently represented as,
\begin{align}
\underset{{{\textbf{f}}_{D}}}{\mathop{\min \text{ }\!\!~\!\!\text{ }}}\,~~~~~~&\textbf{f}_{D}^{H}F_{RF}^{H}{\hat{R}}{{F}_{RF}}{{\textbf{f}}_{D}}\nonumber\\
s.t.~~~~~~~&\textbf{f}_{D}^{H}F_{RF}^{H}\textbf{a}\left( {{\theta }_{d}} \right)=1,
\end{align}
where ${\hat{R}}$ is the estimated array covariance matrix, since a typical information about different signals may not be possible.
\begin{equation}
\hat{R}=~\frac{1}{Q}\underset{q=1}{\overset{Q}{\mathop \sum }}\,{\textbf{x}}\left( q \right){{{\textbf{x}}}^{H}}\left( q \right),
\end{equation}
where $Q$ is the snapshot size. Therefore, we get $\textbf{f}_D$ by solving (21) same like the solution of problem (17) using Lagrange's multiplier as,
\begin{equation}
{{\textbf{f}}_{D}}=~\frac{\hat{R}^{-1}F_{RF}^{H}\textbf{a}\left( {{\theta }_{d}} \right)}{{{(F_{RF}^{H}\textbf{a}\left( {{\theta }_{d}} \right))}^{H}}\hat{R}^{-1}F_{RF}^{H}\textbf{a}\left( {{\theta }_{d}} \right)}
\end{equation}
As $Q$ increases, ${\hat{R}}$ will converge to the true covariance matrix. However, the convergence of the Standard Capon Beamformer (SCB) is so slow. To resolve this problem, a widely used diagonal loading method has been used in order to improve SCB performance. The diagonal loading technique can improve the performance of SCB by adding to the covariance matrix an identity matrix scaled by a real weight called diagonal loading level [44]. Assume $\xi $ is the proposed diagonal loading level, the new digital beamformer vector can be given by,
\begin{equation}
{{\textbf{f}}_{D}}=~\frac{\left( \hat{R}+\xi I \right)^{-1}F_{RF}^{H}\textbf{a}\left( {{\theta }_{d}} \right)}{{{(F_{RF}^{H}\textbf{a}\left( {{\theta }_{d}} \right))}^{H}}\left( \hat{R}+\xi I \right)^{-1}F_{RF}^{H}\textbf{a}\left( {{\theta }_{d}} \right)}
\end{equation}

The diagonal loading level $\xi$ has a considerable effect on the performance of SCB; therefore, several methods have been proposed to optimize the diagonal loading level. One of the most effective and simple methods is the spatial matched filter (SMF) method [45]. The loading level of SMF is given by,
\begin{align}
{{\xi }_{SMF}}&=~\frac{1}{Q}{{\left\| \hat{\textbf{a}}\left( {{\theta }_{d}} \right)X \right\|}^{2}}\nonumber\\
&=~{{\hat{\textbf{a}}}^{H}}\left( {{\theta }_{d}} \right){{\hat{R}}_{x}}\hat{\textbf{a}}\left( {{\theta }_{d}} \right),
\end{align}
where $\hat{\textbf{a}}\left( {{\theta }_{d}} \right)=~\frac{\textbf{a}\left( {{\theta }_{d}} \right)}{\left\| \textbf{a}\left( {{\theta }_{d}} \right) \right\|}$ is the normalized steering vector, and ${{\hat{R}}_{x}}=\frac{1}{Q}\textbf{x}{{\textbf{x}}^{H}}$ is the estimation covariance matrix of the received signal.

Although the above-presented complex weights' methods and other similar methods are very impressive mathematically and fast; however, it requires an expensive receiver making them impractical. Moreover, these algorithms get trapped in local minima.
Consequently, we propose an efficient metaheuristic optimization algorithm in order to optimize the digital beamformer weights using only the phase, where as shown in Eq. (7), digital beamformer vector can be adjusted using amplitude only (i. e., $a_1$,$a_2$,...,$a_K$), phase only (i. e., ${{e}^{j{{\alpha }_{1}}}},\text{ }\!\!~\!\!\text{ }{{e}^{j{{\alpha }_{2}}}},\text{ }\!\!~\!\!\text{ }.\text{ }\!\!~\!\!\text{ }.\text{ }\!\!~\!\!\text{ }.\text{ } \!\!~\!\!\text{ },\text{ }\!\!~\!\!\text{ }{{e}^{j{{\alpha }_{K}}}}$), or complex.

\subsection{Design of the analog weight vectors}
According to the unitary constraint, the cost function given in (16) can further be simplified as,
\begin{align}
\underset{{{F}_{RF}}}{\mathop{minimize}}\,~~~CF=~~&\frac{~\textbf{f}_{D}^{H}F_{RF}^{H}{{A}_{i}}A_{i}^{H}{{F}_{RF}}{{\textbf{f}}_{D}}+~{{\sigma }^{2}}~\textbf{f}_{D}^{H}{{\textbf{f}}_{D}}}{{{P}_{d}}}\nonumber\\
s.t.~~~~~~~~~~~~~~~~&\textbf{f}_{D}^{H}F_{RF}^{H}\textbf{a}\left( {{\theta }_{d}} \right)=1,
\end{align}
where $\textbf{f}_{D}$ is given by (24), therefore, the second term in the numerator and the denominator are constant, thus, the optimization problem in (26) can be reformulated as,
\begin{align}
\underset{{{\theta }}}{\mathop{minimize}}\,~~~&CF~~={{\left| \textbf{f}_{D}^{H}F_{RF}^{H}{{A}_{i}} \right|}^{2}}\nonumber\\
s.t.~~~~~~~&\textbf{f}_{D}^{H}F_{RF}^{H}\textbf{a}\left( {{\theta }_{d}} \right)=1,
\end{align}
where ${{\theta }}$ is the searching angle in the range $\frac{-\pi }{2}~\le {{\theta }}\le \frac{\pi }{2}$, which will be used to construct $F_{RF}$, where, as shown in Eq. (3), the matrix $F_{RF}$ can be built using the corresponding phase of the $mth$ antenna of subarray $l$, ${{\alpha }_{m,l}}$.
\begin{equation}
{\alpha }_{m,l}=~\frac{2\pi }{\lambda }\left( \left( l-1 \right)M+m+1 \right)dsin\theta
\end{equation}
By adjusting the value of $\theta$ in (28), we can minimize the cost function in (27). This can be done using APALS with sufficiently fine grid of points in the above defined range of $\theta$. To perform linear fine searching, we will use small enough searching step size $\Delta \theta $, where the range of $\theta$ is divided into $N_{step}$ sub-periods or points. Therefore, the angle $\theta$ in (28) is selected from the set, $\Theta \text{ }\!\!~\!\!\text{ }\in \text{ }\!\!~\!\!\text{ }\left\{ -\frac{\pi }{2},-\frac{\pi }{2}+\Delta \theta ,\text{ }\!\!~\!\!\text{ }.\text{ }\!\!~\!\!\text{ }.\text{ }\!\!~\!\!\text{ }.\text{ }\!\!~\!\!\text{ }.,\text{ }\!\!~\!\!\text{ }\frac{\pi }{2} \right\}$.

\section{Proposed evolutionary optimization algorithm}\label{sec4}
Evolutionary optimization algorithms are also known as meta-heuristic algorithms such as GA, PSO, and BA algorithms have become recently popular and very efficient that can easily solve many hard optimization problems. In this paper, these algorithms have been proposed to design HAD beamforming by optimizing $f_{D}$ weight vector in order of maximizing the SINR.

\subsection{Bat algorithm}
The BA is inspired from the advanced echolocation capability of bats used to sense distance in order to avoid barrier and detect prey. It is a promising optimization algorithm that characterized by robustness, accuracy, and fast convergence compared to its predecessors such as genetic algorithm and PSO.
In BA, bats fly randomly at position $x_{i}^{t}$ with velocity $v_{i}^{t}$, loudness $A_{i}^{t}$, frequency $f_{i}$ in a range $\left[ {{f}_{min}},~\,{{f}_{max}} \right]$, and the pulse rate of emission $r_{i}^{t}$ in the range of [0, 1].
The new solutions are obtained by updating the positions and velocities for the group of microbats at time step t which can be given by:
\begin{align}\label{xz}
&{{f}_{i}}=~{{f}_{min}}+\left( {{f}_{max}}-{{f}_{min}} \right)\beta\\
&v_{i}^{t}=~~v_{i}^{t-1}+\left( x_{i}^{t}-{{x}_{*}} \right)f\\
&x_{i}^{t}=~~x_{i}^{t-1}+v_{i}^{t},
\end{align}
where $\beta \in \left[ 0,~1 \right]$ is a random vector drawn from a uniform distribution. ${{x}_{*}}$ is the current global best location (solution) which is located after comparing all the solutions among all the N microbats. $f_{min}$ and $f_{max}$ are chosen depending on the domain size of the interested problem. When one solution is chosen from the current best solutions in local search, a new solution for each microbat is produced locally utilizing random walk as:
\begin{equation}\label{xxz}
{{x}_{new}}=~{{x}_{old}}+\varepsilon {{A}^{t}},
\end{equation}
where $\varepsilon \in \left[ 0,~1 \right]$ is a random number, $A^{t}$ is the average loudness of all the microbats at time step $t$. Moreover, as iteration progress, the loudness $A_{i}$ and the rate $r_{i}$ of emission pulse can be updated by,
\begin{equation}\label{xzz}
A_{i}^{t+1}=~~\alpha A_{i}^{t},~~~~r_{i}^{t+1}=~~r_{i}^{0}[1-\text{exp}\left( -\gamma t \right),
\end{equation}
where $\alpha$ and $\gamma$ are constants $(0<\alpha <1,\,\gamma >0)$ which can be tuned experimentally. After the iterations completed, the global best ${{x}_{*}}$ will be found and used as the optimal result [46]. We summarize the basic steps of BA in algorithm 1. \\
\begin{algorithm} [!t]  
\DontPrintSemicolon
\SetAlgoLined
{Input: Initializing the bat population ${{x}_{i}},~{{v}_{i}},~{{f}_{i}},~{{A}_{i}},~{{r}_{i}}$, the number of iterations N, the population size n, bounds limits Lb, Ub, and the desired and interferences angles ${{\theta }_{d}},~{{\theta }_{j}}$.}
\BlankLine
\For{($i< $ Max number of iterations)}{
   Find the current best solutions by updating velocities and locations, and adjusting frequency, Eqs. (29)-(31).\;
  \If{(rand $> r_{i}$)}{
       Select a solution among the best solutions.\;
       Generating a local solution around the selected best solution.\;
      }
  \If{((rand $< A_{i}$)) and $f\left( {{x}_{i}} \right)<f\left( {{x}_{*}} \right)$}{
       Updating new solution.\;
       Increasing $r_{i}$ and reducing $A_{i}$.\;
      }
   Ranking the bats and find the current best ${{x}_{*}}$\;
   $i=i+1$\;
}
{Output: global best, and the beamforming weights.}\;
\caption{proposed Hybrid Beamformer Based on BA}
\end{algorithm}

\subsection{Improved bat algorithm}
BA received a number of improvements attempts in recent years in order to address the algorithm's shortcomings such as unstable results and being trapped in the local minima [41-43]. In this paper, we adopt the novel BA (NBA) proposed by Xian-Bing Meng et al. [42]. In addition, more refinement has been made in order to improve stability by tuning the inertia weight using random variables to help the algorithm easily skip out of the local minima [41]. The NBA algorithm differs from the classic BA in the following points,
\begin{enumerate}
\item In classic BA, the bats search for their food in one habitat, while in NBA they can do that in different habitats.
\item No consideration for the Doppler effects in BA, whilst all bats can compensate for Doppler effects in echoes according to the closeness of their targets.
\item In NBA, the bats have quantum behavior instead of mechanical behavior, so that a bat can access any position in the entire search area with a specific probability.
\end{enumerate}

The bats$'$ habitat selection depends on a stochastic decision, if a uniform random number $R\in \left[ 0,~1 \right]$ is less than the threshold of the selection $P\in \left[ 0,~1 \right]$, bats will choose the quantum behavior instead of mechanical.
In quantum behavior, bats can search in a wide range of habitats. If a food has been found in one of these habitats, others would come to it as soon as discovered. Thus, the positions of the bats can be represented as,
\begin{equation}
x_{ij}^{t+1}=\left\{ \begin{matrix}
   g_{j}^{t}+\theta *\left| m_{j}^{t}-x_{ij}^{t} \right|*ln\left( \frac{1}{{{u}_{ij}}} \right),\,\,if~ran{{d}_{j}}()<P,  \\
   g_{j}^{t}-\theta *\left| m_{j}^{t}-x_{ij}^{t} \right|*ln\left( \frac{1}{{{u}_{ij}}} \right),~\,\,\,\,otherwise,~~~~~  \\
\end{matrix} \right.
\end{equation}
where $x_{ij}^{t}$ is the $N$ bats' positions in a D-dimensional space, $i\in \left[ 1,~.~.~.,~N \right]$, $j\in \left[ 1,~.~.~.,~D \right]$. $\theta$ is the contraction -expansion coefficient, $m_{j}^{t}$ is the mean of the individual's best position in a D-dimensional space at time step $t$, and $u_{ij}$ is uniformly distributed in the range between 0 and 1. On the other side, with mechanical behavior of bats the frequency can change due to the relative motion between the bat and the prey, so as, it must rely on the bats$'$ compensation rates for the Doppler effect. When the bat, prey, or both move away from each other the frequency decreased, and increased in the vice versa case. The new mathematical model can be represented as follows,
\begin{align}\label{zzs}
&{{f}_{i}}=~{{f}_{min}}+\left( {{f}_{max}}-{{f}_{min}} \right)*rand\left( 0,~1 \right)\\
&f_{i,j}^{,}=~\frac{\left( c+v_{i,j}^{t} \right)}{c+v_{i,j}^{t}}*{{f}_{i,j}}*\left( 1+{{C}_{i}}*\frac{\left( g_{j}^{t}-x_{i,j}^{t} \right)}{\left| g_{j}^{t}-x_{i,j}^{t} \right|+\varepsilon } \right)\\
&v_{i,j}^{t+1}=w*v_{i,j}^{t}+(g_{j}^{t}-x_{i,j}^{t})*{{f}_{i,j}}\\
&x_{i,j}^{t+1}=x_{i,j}^{t}+v_{i,j}^{t},
\end{align}
where, $C$ is a positive number represents the compensation rates, $\varepsilon$ is the smallest constant number in computer used to avoid zero-division error, $c$ is the speed of signal in the air ($c=340~m/s$), and $w$ is an inertia weight parameter that can be adjusted in order to improve the stability of the algorithm by quickly jump out of the local minima, which is given by,
\begin{equation}\label{zss}
w=~{{\mu }_{min}}+\left( {{\mu }_{max}}-{{\mu }_{min}} \right)*rand()+\sigma \times randn(),
\end{equation}
where, ${{\mu }_{min}},{{\mu }_{max}}$  are the minimum and maximum factor of the stochastic inertia weight, respectively. $rand(),~\,randn()$ are the random number between 0 and 1, and random number of standard normal distribution, respectively. $\sigma$ is the deviation between the stochastic inertia weight and its mean.
With regard to local search, since it is required that the bat approaching the prey silently as possible, bats would decrease the loudness and increase the rate of the pulse emission. In addition, the loudness created by other bats in the vicinity should be also taken into consideration. The local generation new position for each bat can be represented as follows,
\begin{align}\label{zxz}
&if~(randn\left( 0,~1 \right)>~{{r}_{i}})\\
&x_{i,j}^{t+1}=g_{j}^{t}*\left( 1+randn\left( 0,~~{{\sigma }^{2}} \right) \right)\\
&{{\sigma }^{2}}=\left| A_{i}^{t}-A_{mean}^{t} \right|+\varepsilon,
\end{align}
where $randn(0,{{\sigma }^{2}})$ is a Gaussian distribution with 0 mean and variance ${\sigma}^{2}$.  $A_{mean}^{t}$ is the average loudness of all bats at time step $t$. Based on the above description, the basic steps are summarized in algorithm 2. \\

\begin{algorithm} [!t]
\DontPrintSemicolon
\SetAlgoLined
{Input: Initializing the bat population, the number of iterations N, the population size n, bounds limits Lb, Ub, the desired and interferences angles ${{\theta }_{d}},~{{\theta }_{j}}$, the parameters of original BA, $\alpha ,~\,\gamma ,~\,{{f}_{min}},~~{{f}_{max}},~\,{{A}_{o}},\,~{{r}_{o}}$, the parameters of IBA $P\,,C\,,~\theta \,,~\,G,~\sigma$.}
\BlankLine
\For{($i< $ Max number of iterations)}{
  \eIf {($rand\left( 0,~1 \right)~<~P$)}{
       Generating new solutions using Eq. (34).\;
       }{   
       Generating new solutions using Eqs. (35)-(39).\;
      }
  \If{(rand $> r_{i}$)}{
      Generating a local solution around the selected best solution using Eqs. (40)-(42).\;
      }
      Select a solution among the best solutions.\;
      Updating new solutions, the loudness and pulse emission rate using Eq. (33).\;
      Ranking the bats and find the current best ${{g}^{t}}$.\;
  \If{the current best does not improve in G time step}{
      Re-initialize the loudness $A_{i}$ and set temporary pulse rates $r_{i}$ as a uniform random number between [0.85, 0.9].\;
      }
      $i=i+1$ \;
      }
{Output: global best, and the beamforming weights.} \;
\caption{proposed Hybrid Beamformer Based on Improved-BA}
\end{algorithm}

\begin{figure}[t!]
\centerline{\includegraphics[width=8.8cm, height=6.4cm]{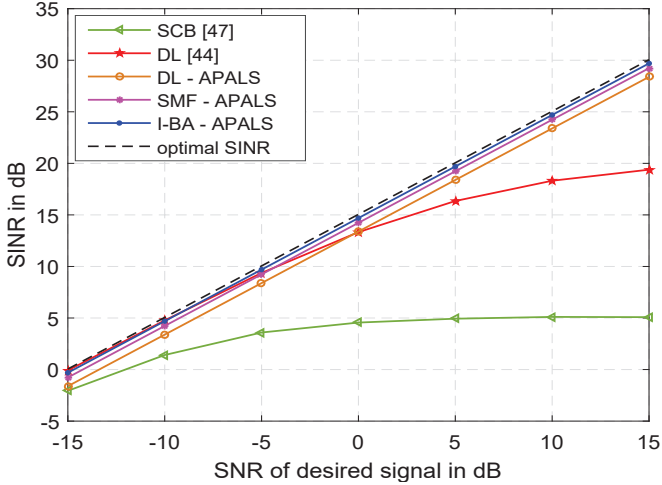}}
\caption{SINR comparison of DL techniques and proposed Improved-BA in the absence of mismatch, N=32, and Q=128, the number of population = 40 for proposed algorithm}
\label{fig:234}
\end{figure}

\begin{table*} [t]
\caption{Parameters used for different algorithms}
\centering
\label{table1}
\begin{tabular} {p{70pt}p{400pt}}   
\hline \\ [-1.5ex]
{Algorithm}& 
{parameters} \\    [1.1ex]
\hline \\ [-1.5ex]

{PSO} & ${{f}_{min}}=0,\,{{f}_{max}}=2,\,w\in \left[ 0.4,~0.9 \right],\,{{C}_{1}}=~0.5,\,{{C}_{2}}=~0.5$.\\  [0.5ex]
{BA} &$\alpha =\gamma =0.9,\,{{f}_{min}}=0,\,{{f}_{max}}=2,\,{{A}_{o}}\in \left[ 0,\,\,2 \right],\,{{r}_{o}}\in \left[ 0,\,1 \right]$. \\ [0.5ex]
{Improved-BA} &$\alpha =\gamma =0.9,\,{{f}_{min}}=0,\,{{f}_{max}}=1.5,\,{{A}_{o}}\in \left[ 0,\,\,2 \right],\,{{r}_{o}}\in \left[ 0,\,1 \right],\,G=2,\,P\in \left[ 0.5,~0.9 \right],\,C\in \left[ 0.1,~0.9 \right],\,\theta \in \left[ 0.5,~1 \right],\,$ \\
             &${{\mu }_{min}}=0.4,\,{{\mu }_{max}}=0.9,\,\sigma =0.2$. \\  [1ex]
\hline
\end{tabular}
\end{table*}

\section{Simulation results}
\label{sec:5}
\begin{figure}[t!]
\centerline{\includegraphics[width=8.4cm, height=12.0cm]{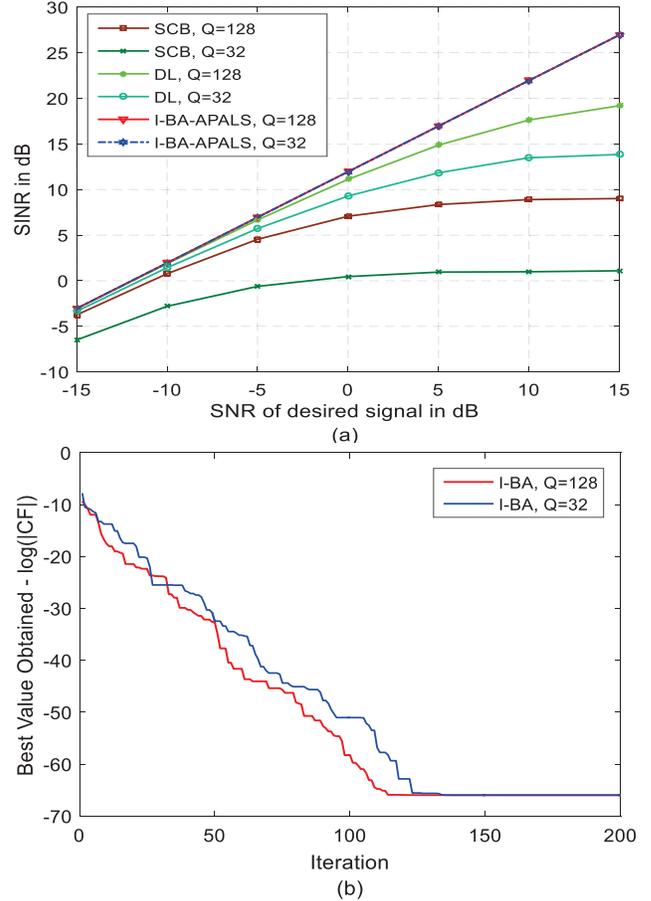}}
\caption{The effects of snapshots size on the output SINR in the absence of mismatch, $N=16$, a) SINR versus $SNR_d$, b) Convergence characteristic curves, the number of population = 40}
\label{fig:3}
\end{figure}

In this section, the performance of Improved-BA (I-BA), DL, and DL-SMF in order to achieve maximum SINR and nulls placement in the desired directions using partial - connected HAD receive beamformer has been evaluated and compared with other efficient fully digital conventional optimization algorithms, namely, SCB [47], DL technique [44]. The simulation parameters used by the proposed algorithm and other evolutionary algorithms were presented in Table 1. Other parameters set as: diagonal loading level $\xi =30$, half-wavelength antenna spacing between adjacent elements, the antenna array receives the desired signal from angle ${{\theta }_{d}}={{0}^{o}}$, and two interference signals arriving from angles ${{\theta }_{k}}={{60}^{o}},~\,-{{30}^{o}}$. The SNR of the interference signals is assumed to be 15.

\begin{table*} [t]
\caption{Comparison the results of nulls depth for different algorithms, when the number of antenna array elements equal 16, and Q=128}
\centering
\label{table3}
\begin{tabular} {lcccc} 
\hline  \\ [-1.5ex]
 \multirow{2}{*}{Algorithm}  & \multicolumn{2}{c}{Nulls Depth in dB} & \multirow{2}{*}{Optimized
angle $(\theta)$} & \multirow{2}{*}{Optimized digital beamformer phases $({{\alpha }_{1}},~{{\alpha }_{2}},~{{\alpha }_{3}},~{{\alpha }_{4}})$} \\ 
                    & ${-3}{{0}^{o}}$  & ${6}{{0}^{o}}$  \\ 

\hline \\ [-1.5ex]
 {SCB [47]}      & -18.2325  & -19.1922     & $-$        &  $-$   \\
 {DL [44]}       & -23.2880  & -21.2294     & $-$        &  $-$   \\
 {DL-APALS}         & -36.3780  & -19.9926  & 1.0367e-04 &  $-$   \\
 {SMF-APALS}        & -36.3803  & -21.9080  & 1.0367e-04 &  $-$   \\
 {I-BA-APALS}      & -36.3947  & -37.8852   & 1.0367e-04 &  0.2285, 0.2618, 0.0029, 0.0362  \\  
\hline
\end{tabular}
\end{table*}

\begin{table*} [t]
\caption{Comparison the results of nulls depth for different algorithms, when the number of antenna array elements equal 32, and Q=128}
\centering
\label{table4}
\begin{tabular} {lcccc} 
\hline \\ [-1.5ex]
 \multirow{2}{*}{Algorithm}  & \multicolumn{2}{c}{Nulls Depth in dB} & \multirow{2}{*}{Optimized
angle $(\theta)$} & \multirow{2}{*}{Optimized digital beamformer phases $({{\alpha }_{1}},~{{\alpha }_{2}},~{{\alpha }_{3}},~{{\alpha }_{4}})$} \\ 
                     & ${-3}{{0}^{o}}$  & ${6}{{0}^{o}}$  \\ 
 \hline \\ [-1.5ex]
 {SCB [47]}      & -19.3760  & -24.7878     & $-$        &  $-$   \\
 {DL [44]}       & -26.2609  & -22.0527     & $-$        &  $-$   \\
 {DL-APALS}         & -36.3777  & -20.0879  & 1.0367e-04 &  $-$   \\
 {SMF-APALS}        & -36.3835  & -24.0793  & 1.0367e-04 &  $-$   \\
 {I-BA-APALS}      & -36.3789  & -37.9193   & 1.0367e-04 &  0.2346, 0.0091, 0.2701, 0.0445  \\  [1ex]
\hline
\end{tabular}
\end{table*}

\begin{figure}[!t]
\centerline{\includegraphics[width=8.4cm, height=12.0cm]{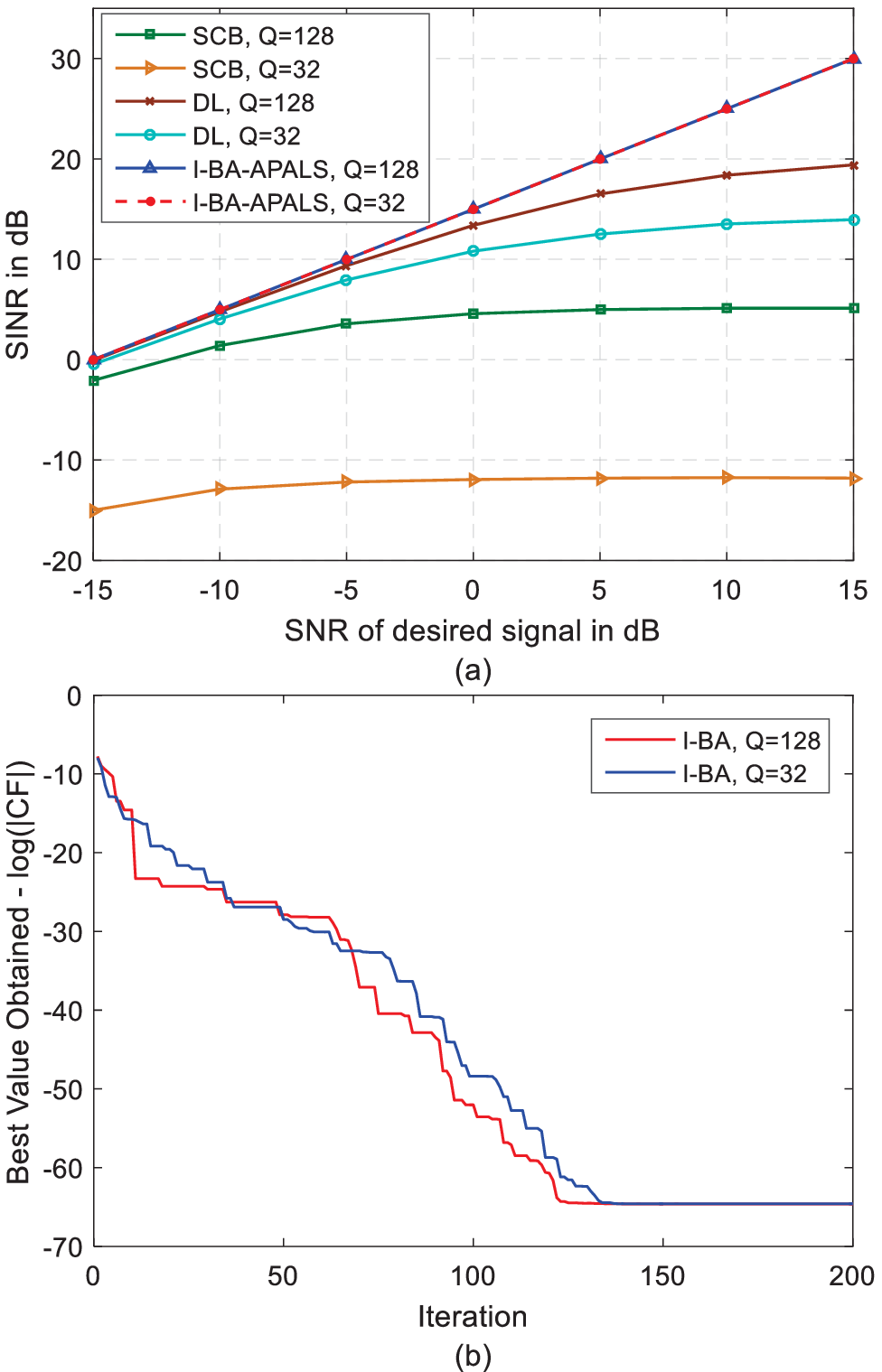}}
\caption{The effects of snapshots size on the output SINR in the absence of mismatch, $N=32$, a) SINR versus $SNR_d$, b) Convergence characteristic curves, the number of population = 40}
\label{fig:4}
\end{figure}

\begin{figure}[!t]
\centerline{\includegraphics[width=8.4cm,height=6cm]{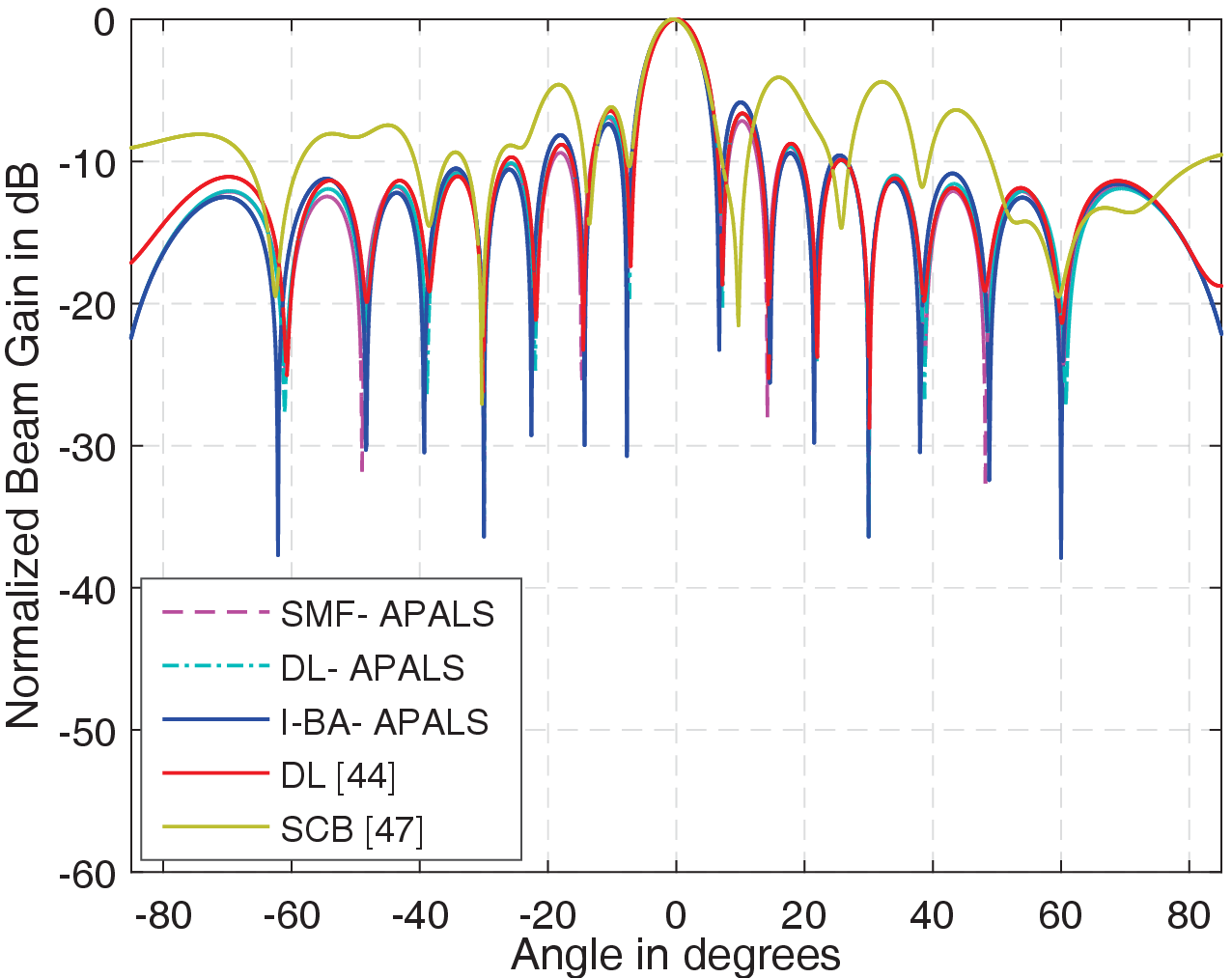}}
\caption{Radiation pattern with nulls placement at ${{-30}^{o}}$, and ${{60}^{o}}$, the number of antenna array elements equal 16, Q=128}
\label{fig:5}
\end{figure}

\begin{figure}[!t]
\centerline{\includegraphics[width=8.4cm,height=6cm]{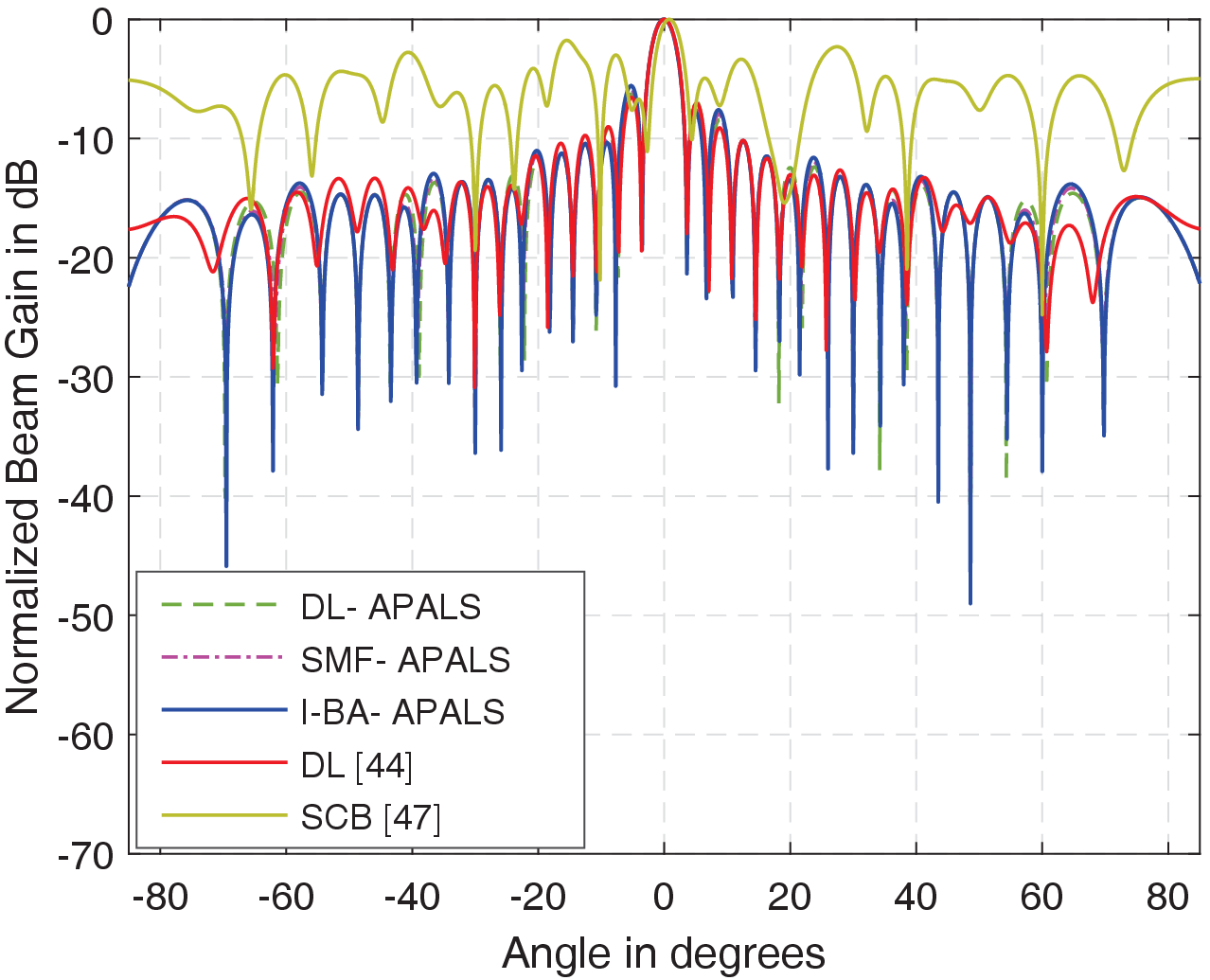}}
\caption{Radiation pattern with nulls placement at ${{-30}^{o}}$, and ${{60}^{o}}$, the number of antenna array elements equal 32, Q=128}
\label{fig:6}
\end{figure}

\begin{figure*}[t]
\centerline{\includegraphics[width= 17.2cm,height=12.8cm]{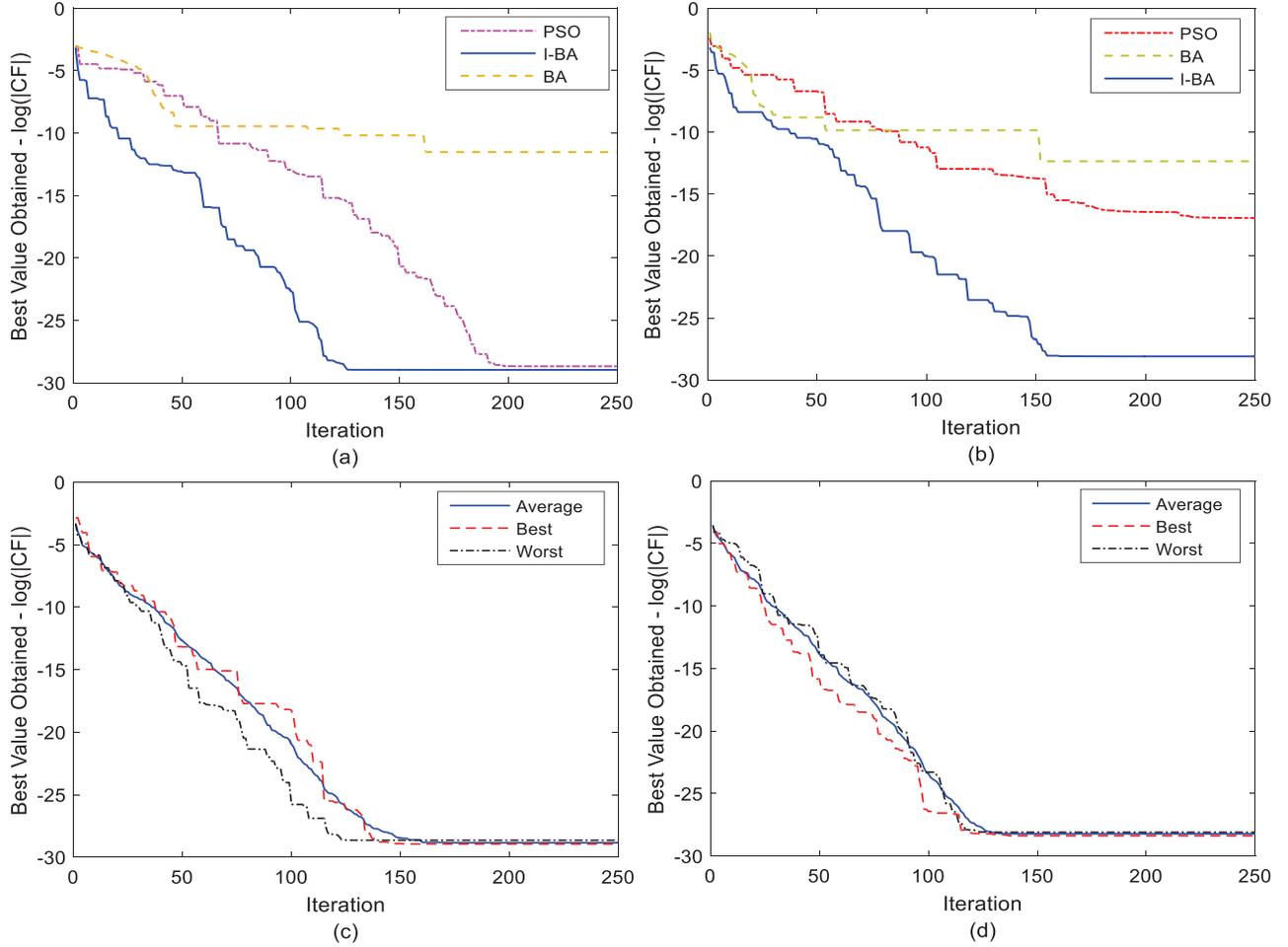}}
\caption{Convergence characteristics curves for varying $SNR_d$, $N$, and snapshots. the population size equal 40 for all algorithms and cases, (a) $N=16$, snapshots = 128, $SNR_d = 0$, (b) N=32, snapshots = 200, $SNR_d = -15$ , (c) $N=16$, snapshots = 128, $SNR_i = 30$, $SNR_d = -15$, (d) N=32, snapshots = 200, $SNR_i = 30$, $SNR_d = 0$}
\label{fig:8}
\end{figure*}

\begin{figure}[!t]
\centerline{\includegraphics[width=8.4cm,height=6cm]{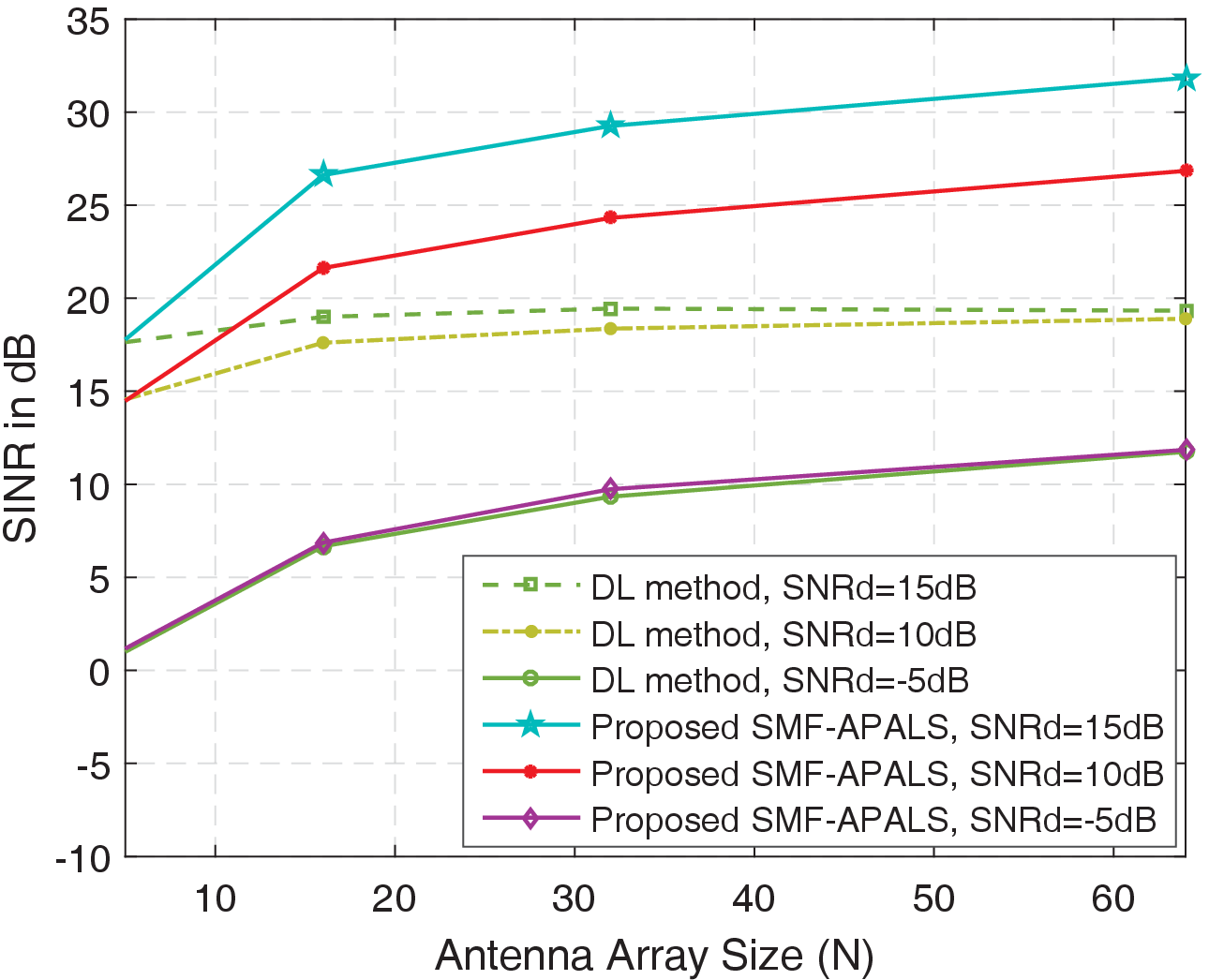}}
\caption{The effects of varying the number of antennas into SINR when the SNR of interference signals are fixed ($SNR_i$  =15 dB) for different $SNR_d$, snapshots =128}
\label{fig:7}
\end{figure}

Fig. 2 presents the output SINR versus input SNR of the desired signal, in this examination, $Q$ is set to be 128, and $N$ to be 32. Fig. 2 compares RAB techniques that have been used to optimize digital beamforming vector and the proposed evolutionary optimization algorithm that employed to optimize the digital beamforming vector by controlling the phase only. Observing Fig. 2, we notice that our proposed evolutionary optimization technique has better performance compared to all other techniques. The hybrid I-BA - APALS has a best performance which is highly close to the optimal followed by SMF-APALS, which has a very close performance, while the traditional SCB and DL methods have very poor performance for higher $(SNR_d)$, especially the SCB method.

Fig. 3 (a) and Fig. 4 (a) present the curves of SINR versus SNR of the desired signal $(SNR_d)$ for snapshots size 32 and 128 using our proposed Improved-BA (I-BA) algorithm and compared the results with DL technique and SCB when the antenna array size equal 16 and 32, respectively. The SNR of interference signals $(SNR_i)$ is chosen to be 15dB. Observing these figures, we find that the SCB method has a very low performance compared with our proposed algorithm and DL technique as expected. The significant degradation of SCB shown in Fig. 4 (a) is because of the growing difference between the estimated array covariance matrix $\hat{R}$ and the true covariance matrix as N increases. For the $(SNR_d)$ less than -5dB, we notice that the DL technique and our proposed algorithm have almost the same performance, however, for higher values of $(SNR_d)$ the performance of our proposed algorithm is significantly better than DL technique. Therefore, our proposed algorithm has a better ability to minimize interference. On the other hand, Fig. 3 (a) and Fig. 4 (a) demonstrate the effect of the number of snapshots to the performance of different algorithms, where the performance is improved considerably as the number of snapshots increased for SCB and DL techniques, whereas, there is very little effect on the performance of our proposed algorithm with significantly lower snapshots size.

The optimized beamforming gains enjoined with two nulls at $-{30}^{o}$, and ${60}^{o}$ for a number of antenna array elements 16 and 32 have been given in Fig. 5 and Fig. 6, respectively. The $SNR_d$ and $SNR_i$ are chosen to be -5dB and 15dB, respectively. The hybrid methods I-BA-APALS, SMF-APALS, and DL-APALS, in addition to the traditional techniques SCB [47] and DL [44], are used to synthesize a linear array. It can be seen obviously that the proposed I-BA-APALS algorithm has better ability to mitigate the beam gain by at least -36dB at the predefined locations of interference signals compared to other methods as shown in Table 2 and Table 3.

\begin{figure}[!t]
\centerline{\includegraphics[width=8.4cm,height=6cm]{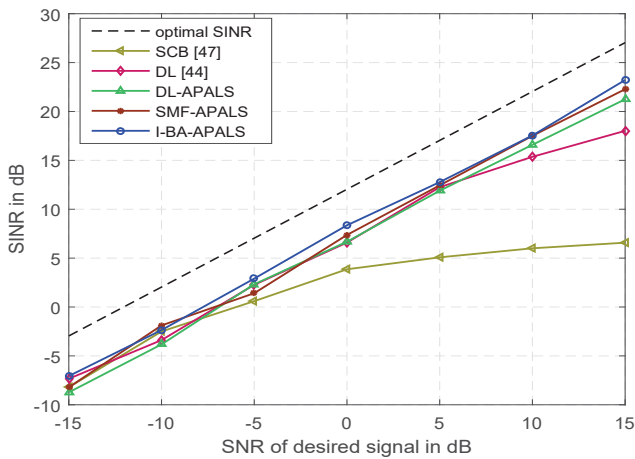}}
\caption{Output SINR versus input SNR of the desired signal, $N=16$, and Q=200, in the presence of DOA mismatches}
\label{fig:9}
\end{figure}

\paragraph{Convergence characteristics}
Fig. 7 shows a comparative convergence characteristic graphs obtained using BA, PSO, and proposed algorithm. All algorithms are used to solve the objective function given in (27) with a different number of antennas, snapshot size, $SNR_i$, and $SNR_d$. In Fig. 7 (a) and Fig. 7 (b), the number of antenna array elements are set to be 16 and 32, whereas, the snapshots size are set to be 128 and 200, respectively. The $SNR_i$ is chosen to be 15, and $SNR_d$ is set to be 0 and -15, respectively. Observing these curves, it is shown that the BA gets local optimal solution early in both figures; however, our proposed algorithm has better ability to jump from local minima in both cases. Although PSO has close optimal solution to I-BA in Fig. 7 (a), however, it has poor performance for higher antenna array size and weak $SNR_d$. The comparison of average convergence curves of the proposed algorithm is illustrated in Fig. 7 (c) and  Fig. 7 (d) for 20 runs, which shows a quite good stability.

Fig. 8 illustrates the effect of antenna array size on the performance of hybrid beamformer proposed by us. From Fig. 8, it is shown that the performance of the proposed beamformer based on SMF-APALS is gradually improved with increasing the number of antennas. As noticed from the curves, the different values of the received SNR of the desired signal almost have the same impact on the performance improvement for our proposed algorithm. On the other hand, while our proposed algorithm showed important improvement on the performance as antenna array size increased, DL method has no significant impact on the performance for higher values of $SNR_d$.

\paragraph{The impact of DOA mismatches}
Finally, Fig. 9 examines the effect of DOA mismatch into the performance of proposed robust adaptive methods, where, the number of antenna array elements and snapshot size are set to be 16 and 200, respectively. The maximum estimation DOA angle mismatch is chosen to be $3^o$. Fig. 9 further compares the performance of the classic SCB, and DL methods with the hybrid, I-BA-APALS, DL- APALS, and SMF- APALS proposed techniques in the presence of DOA mismatch, where the proposed I-BA-APALS showed better robustness performance to the DOA mismatch followed by SMF- APALS with very close performance. This because the proposed I-BA-APALS has good flexibility, therefore, it has a less impact by the DOA mismatch, the number of antenna array elements, and snapshot size.

\section{Conclusion}\label{sec6}
In this paper, we have proposed a hybrid beamforming system based on three hybrid adaptive beamforming techniques, namely, DL-APALS, SMF-APALS, and I-BA-APALS with the objective of maximizing SINR. The first two methods using only a linear searching to obtain the optimal solution, where, the optimum digital beamformer vector is obtained by closed-form solution, and the linear searching is employed to optimize the analog beamformer vectors. In the last hybrid scheme, we further proposed an efficient nature-inspired optimization technique, that is, I-BA with aim of optimizing the digital beamforming vector, which gave better global optima, convergence speed, and stability performance as compared to BA, and PSO. With the aid of simulation and analysis, we find that the performance of the traditional adaptive beamformers, i.e., SCB, and DL techniques have serious degradation when the input SNR of the desired signal is large. By combining the DL, SMF, and I-BA methods with linear searching scheme to optimize the total beamforming vector, we got a better performance by I-BA-APALS in terms of output SINR, nulls depth, and robustness against DOA mismatch followed by SMF-APALS. On the other hand, since the I-BA-APALS depends on the phase-only as a controlling parameter; resulting in an inexpensive receiver making it convenient for practical implementation. This makes our proposed beamformer appropriate for future many applications that are likely to be susceptible to interference such as future 5G wireless cellular communication systems, UAVs, and intelligent transportation.

\section{Acknowledgments}\label{sec7}
This work was supported in part by the National Natural Science Foundation of China (Nos. 61771244, 61701234, 61501238, 61702258, 61472190, and 61271230), in part by the Open Research Fund of National Key Laboratory of Electromagnetic Environment, China Research Institute of Radiowave Propagation (No. 201500013), in part by the Jiangsu Provincial Science Foundation under Project BK20150786, in part by the Specially Appointed Professor Program in Jiangsu Province, 2015, in part by the Fundamental Research Funds for the Central Universities under Grant 30916011205, and in part by the open research fund of National Mobile Communications Research Laboratory, Southeast University, China (Nos. 2017D04 and 2013D02).

\end{document}